\documentclass[aps,prl,twocolumn,longbibliography
,groupedaddress]{revtex4-1}

\usepackage{amsmath}
\usepackage{graphicx}

\newcommand{\be}{\begin{equation}}
\newcommand{\ee}{\end{equation}}
\newcommand{\bse}{\begin{subequations}}
\newcommand{\ese}{\end{subequations}}
\newcommand{\ba}{\begin{eqnarray}}
\newcommand{\ea}{\end{eqnarray}}
\newcommand{\bea}{\begin{eqnarray}}
\newcommand{\eea}{\end{eqnarray}}
\newcommand{\nn}{\nonumber \\}

\newcommand{\g}{g} 
\newcommand{\mn}{{\mu\nu}}

\newcommand{\gt}{\tilde{g}} 
\newcommand{\gB}{g^{\mathrm{(B)}}} 

\newcommand{\x}{a} 
\newcommand{\xt}{\tilde{a}} 
\newcommand{\y}{b} 
\newcommand{\yt}{\tilde{b}} 
\newcommand{\Z}{c} 
\newcommand{\Zt}{\tilde{c}} 

\newcommand{\tit}{\tilde{t}} 

\begin{document}


\title{Hydrodynamic attractor of a hybrid viscous fluid in Bjorken flow}


\author{Toshali Mitra}
\email[]{toshalim@imsc.res.in}
\affiliation{The Institute of Mathematical Sciences, Chennai 600113, India}

\author{Sukrut Mondkar}
\email[]{sukrutmondkar@gmail.com}
\affiliation{Department of Physics, Indian Institute of Technology Madras,\\ Chennai 600036, India}

\author{Ayan Mukhopadhyay}
\email[]{ayan@physics.iitm.ac.in}
\affiliation{Department of Physics, Indian Institute of Technology Madras,\\ Chennai 600036, India}

\author{Anton Rebhan}
\email[]{anton.rebhan@tuwien.ac.at}
\affiliation{Institut f\"{u}r Theoretische Physik, Technische Universit\"{a}t Wien,\\
Wiedner Hauptstr.~8-10, A-1040 Vienna, Austria}

\author{Alexander Soloviev}
\email[]{alexander.soloviev@stonybrook.edu}
\affiliation{Department of Physics and Astronomy, Stony Brook University, \\Stony Brook, New York 11794, USA}

\date{\today}

\begin{abstract}
The nonequilibrium evolution in a boost-invariant Bjorken flow of
a hybrid viscous fluid model containing two interacting components with different viscosities, such that they represent
strongly and weakly self-coupled sectors, is shown to be characterized by a hydrodynamic attractor
which has an early-time behavior that is reminiscent of the so-called
bottom-up thermalization scenario in heavy-ion collisions.
The hydrodynamization times for the two sectors can differ strongly, with details depending
on the curve realized on the two-dimensional attractor surface, which might account for different scenarios for
small and large systems in nuclear collisions.
The total system behaves like a single viscous fluid with a dynamically determined effective shear viscosity.
\end{abstract}


\maketitle

\section{Introduction}
Hydrodynamic models have enjoyed 
an astounding success in describing the collective flow 
of low-$p_T$ hadrons in heavy-ion collisions \cite{Heinz:2013th,Gale:2013da,Florkowski:2017olj,Romatschke:2017ejr,Berges:2020fwq}. These models require hydrodynamics to be initialized at $\lesssim 1$ fm/c after the collisions when the system is far away from equilibrium and with a very low ratio of shear-viscosity to entropy density ($\eta/s \lesssim 0.2$). The studies of heavy-ion collisions in strongly interacting gauge theories using holographic methods \cite{Chesler:2010bi,PhysRevLett.108.201602} were the first to provide theoretical insights not only into the low ratio of shear-viscosity to entropy density \cite{Policastro:2001yc,Romatschke:2017ejr} but also into the applicability of hydrodynamics itself in far away from equilibrium conditions. In particular, it was understood that the hydrodynamic expansion (which generically has zero radius of convergence) needs to be resummed to all orders in derivatives in order to generate a causal evolution which can be matched with the evolution in the exact microscopic theory for an arbitrary initial condition \cite{Heller:2013fn,Heller:2015dha,Buchel:2016cbj}. 
Furthermore, the resummed hydrodynamic expansion can be expressed as an appropriate \textit{trans-series} from which one could also extract the quasi-normal type relaxation modes of the system.

Subsequently, it was established that a wide variety of other phenomenological approaches, such as M\"uller-Israel-Stewart (MIS)
\cite{Muller:1967zza,Israel:1979wp}
and related versions of extended hydrodynamics \cite {Heller:2015dha,Aniceto:2015mto,Strickland:2017kux}, and kinetic theory \cite{Romatschke:2017vte,Strickland:2017kux,Strickland:2018ayk,Almaalol:2020rnu,Denicol:2019lio} also demonstrated the existence of the \textit{hydrodynamic attractor}, which is the evolution obtained from resumming the hydrodynamic expansion to all orders and to which the system approaches rather quickly for any initial condition. The trans-series has been also examined in such contexts, notably in \cite{Heller:2015dha,Basar:2015ava,Aniceto:2015mto,Heller:2016rtz,Heller:2018qvh}. The ubiquitous presence of the hydrodynamic attractor provides a stronger foundation for hydrodynamics as a causal and consistent effective theory that may be applied generally even when the system is yet to achieve local equilibration. The approach to the hydrodynamic attractor provides a more precise meaning to \emph{hydrodynamization} \cite{PhysRevLett.108.201602,Attems:2017zam} of the system, referring to the 
feature that the energy-momentum tensor and conserved currents can be described by an effective hydrodynamic theory very accurately after a time that is commensurate with a microscopic time-scale for any arbitrary initial data (see \cite{Heller:2016gbp} for a nice discussion).
Strong coupling leads to more rapid hydrodynamization \cite{Heller:2015dha} with significant qualitative differences from weak-coupling scenarios \cite{PhysRevLett.124.102301}.

A key feature of quantum chromodynamics (QCD) is asymptotic
freedom, which in ultrarelativistic heavy ion collisions
makes it possible to describe
hard quasiparticle degrees of freedom to some extent
by perturbative QCD (in terms of the glasma effective theory \cite{Gelis:2010nm})
and kinetic theory. However, the gluons produced
in their interactions have a coupling that is the stronger the smaller
their momenta are, so that strong-coupling methods are needed for the
description of the developing bath of soft gluons. 

In the so-called bottom-up thermalization scenario initially proposed
in \cite{Baier:2000sb} and refined in \cite{Kurkela:2011ub}, isotropization
and eventually thermalization results from the build-up of a thermal
bath of soft gluons. This is in contrast to the completely strong-coupling
picture provided by gauge/gravity duality, where thermalization is
instead top-down \cite{Balasubramanian:2011ur}. 
In \cite{Steineder:2012si, Stricker:2013lma} it was attempted to find a transition
from top-down to bottom-up scenarios as the infinite coupling limit
of the standard AdS/CFT correspondence is relaxed through higher-curvature
corrections \cite{Waeber:2015oka}. However, since the quark-gluon medium produced
in heavy ion collisions involves more 
strongly and more weakly coupled sectors
simultaneously, a hybrid approach which combines weak and strong
coupling features may be required.

To this end,
in \cite{Iancu:2014ava,Mukhopadhyay:2015smb,Ecker:2018ucc} the semi-holographic
approach originally developed by \cite{Faulkner:2010tq, Mukhopadhyay:2013dqa} in the context of non-Fermi liquids
was utilized for combining the weak-coupling glasma framework \cite{Gelis:2010nm}
with a holographic AdS/CFT description of the infrared sector of soft gluons \footnote{For another
hybrid approach see \cite{Casalderrey-Solana:2014bpa} which was aimed at
the description of the energy loss of hard jets moving through a strongly coupled medium.}.
In more general terms, a hybrid two-fluid system with couplings analogous to those
used in the semi-holographic approach was introduced and studied in
\cite{Kurkela:2018dku} in thermodynamic equilibrium.

In this Letter we present the first results of a study of the nonequilibrium
evolution of the hybrid fluid model of \cite{Kurkela:2018dku} in a boost-invariant
longitudinal expansion known as Bjorken flow \cite{Bjorken:1982qr}. 
The two components are assumed to have a conformal equation of state,
but different amounts of shear viscosity and relaxation times, chosen so that
one component corresponds to the infinite coupling limit in AdS/CFT \cite{Baier:2007ix} and the
other has significant higher shear viscosity corresponding to a less strongly coupled fluid.
The two fluids can exchange energy and momentum through mutual deformations of an effective metric
that the respective subsystems are living in, while the resulting total energy-momentum tensor
of the full system (which actually lives in Minkowski space) is conserved, but with
nonvanishing trace caused by the interactions.

The nonequilibrium evolution of the full system turns out to involve a hydrodynamic attractor,
which defines a two-dimensional hypersurface in the four-dimensional \footnote{The symmetries
of the effectively 1+1-dimensional Bjorken flow 
admit different longitudinal and transverse pressures for each of the two (conformal) subsystems.}
phase space spanned
by the degrees of freedom of the subsystems. Generically, we find that the attractor
requires an initial evolution of the distribution
of energy in the subsystems that is reminiscent of the bottom-up thermalization scenario, 
where the energy is dominated by the more viscous (weakly self-coupled) sector at early times
and rapidly shared with the less viscous (strong-coupling) sector.

Subsequently, 
significant differences in the hydrodynamization times of the subsystems
are observed, after which
the energy-momentum tensor of the full system can be described hydrodynamically with
a specific viscosity determined by which curve on the attractor hypersurface the system follows.

We want to emphasize, however, that although our model here is inspired from semi-holography, it does not capture all the complexities of the latter. We mainly borrow the idea of democratic couplings between the effective descriptions of the perturbative and non-perturbative sectors, which 
have been shown to pass stringent consistency checks with the principles of thermodynamics, statistical mechanics and Wilsonian renormalization group flow \cite{Kurkela:2018dku,Banerjee:2017ozx}. While a semi-holographic
description involving a similar coupling between a kinetic sector and a dynamical black hole will exhibit irreversible energy transfer to the soft sector on longer time scales, as seen in \cite{Ecker:2018ucc}, in our two-fluid setup there is no such mechanism. 
Our model can therefore capture only certain aspects of the evolution described by a semi-holographic setup
(also the description of the early stages of the weakly coupled sector as a fluid is a drastic simplification).
A noteworthy feature of our model is that the full system can still be described hydrodynamically at late times even when
local equilibration cannot happen due to a lack of sufficient time for mutual interactions between its components.

Besides its intrinsic interest as a tractable two-fluid model with hydrodynamic attractor,
which to our knowledge is the first of its kind with many possibilities of generalizations,
our model is of specific interest to quark-gluon plasma physics, where the interplay of hard and soft
sectors of the dynamics can be modelled in a new way. 
While it is clear that it cannot capture the specific interactions described by QCD, it
might reveal features of the nonequilibrium dynamics of QCD that have a more general character.
For instance, the observed
dependence of the ratio of hydrodynamization times 
on the extra parameters of the attractor surface
might account for different scenarios for small and large systems in nuclear collisions.

\section{Hybrid fluid model with Bjorken flow}
In \cite{Kurkela:2018dku} a minimal coupling of the energy-momentum tensors of two subsystems
was introduced in the following, purely geometrical way,
inspired by the semi-holographic setup of \cite{Iancu:2014ava,Mukhopadhyay:2015smb,Banerjee:2017ozx,Ecker:2018ucc}.
The combined system is defined on one and the same spacetime with metric tensor $\gB_\mn$, but
the two energy-momentum tensors of the subsystems, denoted
by $t^\mn$ and $\tit^\mn$, are assumed to be covariantly conserved only with respect
to \emph{effective} metric tensors $g_\mn$ and $\gt_\mn$, which differ from $\gB_\mn$
and are determined locally by the respective
other subsystem according to
\bea\label{couplingeqs}
g_\mn&=&\gB_\mn+\frac{\sqrt{-\gt}}{\sqrt{-\gB}}\Bigl{[}\gamma \tilde{t}^{\alpha\beta}\gB_{\alpha\mu}\gB_{\beta\nu}+\gamma^\prime {\rm tr}(\tilde{t}\cdot \gB) \gB_\mn \Bigr{]},\nn
\gt_\mn&=&\gB_\mn+\frac{\sqrt{-g}}{\sqrt{-\gB}}\Bigl{[}\gamma t^{\alpha\beta}\gB_{\alpha\mu}\gB_{\beta\nu}+\gamma^\prime {\rm tr}(t\cdot \gB) \gB_\mn \Bigr{]},\nn
\eea
with two coupling constants $\gamma,\gamma'\equiv-r\gamma$ with mass dimension $-4$. 
(We need $\gamma>0$
in order that the dynamics of the subsystems
remains causal with respect to the physical background metric, and $r>1$ for UV completeness \cite{Kurkela:2018dku}.)
These coupling rules ensure that the full system has a conserved energy-momentum tensor in the physical background, $\nabla^\mathrm{(B)}_\mu T^\mu_{\phantom{\mu}\nu} = 0$, with
\begin{eqnarray}\label{fullT}
T^\mu_{\phantom{\mu}\nu} &=& \frac{1}{2}[(t^\mu_{\phantom{\mu}\nu}+t_\nu^{\phantom{\nu}\mu})\frac{\sqrt{-g}}{\sqrt{-\gB}} +
(\tit^\mu_{\phantom{\mu}\nu}+\tit_\nu^{\phantom{\nu}\mu})\frac{\sqrt{-{\tilde{g}}}}{\sqrt{-\gB}}]\nn 
&+& \Delta K \delta_\nu^\mu =: T^\mu_{1\;\nu}(\mathcal{E}_1,\mathcal{P}_1)+T^\mu_{2\;\nu}(\mathcal{E}_2,\mathcal{P}_2)
+T^\mu_{\phantom{\mu}\nu,\mathrm{int}}\quad
\end{eqnarray}
where
\bea
\Delta K = &-\frac{\gamma}{2}\Big(t^{\rho \alpha} \frac{\sqrt{-{g}}}{\sqrt{-{\gB}}}\Big)  \gB_{\alpha \beta} \Big(\tit^{\beta \sigma} \frac{\sqrt{-{\tilde{g}}}}{\sqrt{-{\gB}}} \Big) \gB_{\sigma \rho} \nonumber \\ 
&-\frac{\tilde{\gamma}}{2}\Big(t^{\alpha \beta}  \gB_{\alpha \beta} \frac{\sqrt{-{g}}}{\sqrt{-{\gB}}}\Big) \Big(\tit^{\sigma \rho}\gB_{\sigma \rho} \frac{\sqrt{-{\tilde{g}}}}{\sqrt{-{\gB}}} \Big).
\eea

In the Bjorken-flow case in flat Minkowski spacetime, 
with boost-invariant transversely homogeneous expansion in the $z$-direction,
the background metric will be given in Milne coordinates $(\tau,x,y,\zeta)$ where
$\gB_\mn=\text{diag}(-1,1,1,\tau^2)$ with proper time $\tau=\sqrt{t^2-z^2}$ and spacetime rapidity $\zeta=\tanh^{-1}(z/t)$.
In this situation, a boost-invariant ansatz for the effective metric tensors of the subsystems is given by six scalar
function of $\tau$ according to
\bea
\g_\mn(\tau)&=\text{diag}(-\x^2,\y^2,\y^2,\Z^2),\\
\gt_\mn(\tau)&=\text{diag}(-\xt^2,\yt^2,\yt^2,\Zt^2),
\eea
and the coupling equations (\ref{couplingeqs}) provide six nonlinear algebraic equations to solve for them.

The equations of state of the subsystems are assumed to be conformal, $\epsilon=3P$, $\tilde\epsilon=3\tilde P$, so
that the ansatz for the energy-momentum tensors of the subsystem reads
\bea
t^\mn&=&\text{diag}\Big{(}\frac{\varepsilon}{\x^2},\frac{P_\perp}{\y^2},\frac{P_\perp}{\y^2},\frac{P_L}{\Z^2} \Big{)},\nonumber\\
&=&\text{diag}\Big{(}\frac{\varepsilon}{\x^2},\frac{P}{\y^2},\frac{P}{\y^2},\frac{P}{\Z^2} \Big{)}+\pi^\mn,\\
\pi^\mn&=&\text{diag}(0,\frac{\phi}{2\y^2},\frac{\phi}{2\y^2},-\frac{\phi}{\Z^2}),
\eea
and analogously for $\tit^\mn$. 
Notice, however, that the full system \eqref{fullT} is not conformal, but has a nonvanishing trace of the energy-momentum
tensor produced by the interactions of the two subsystems, which involve the dimensionful coupling $\gamma$, 
whose scale could be set by the saturation momentum $Q_s$ of the glasma effective theory \cite{Gelis:2010nm}.

Covariant conservation with respect to the effective metric (but not with respect to the physical background metric),
$\nabla_\mu t^\mn=0$ yields
\be\label{WIeq}
\partial_\tau \varepsilon +\varepsilon \partial_\tau \log( \y^{8/3}\Z^{4/3})+\phi\partial_\tau \log{\y/\Z}=0.\\
\ee
The dynamical system is closed by assuming relaxation of $\pi^\mn$ by MIS equations
with shear viscosities $\eta,\tilde\eta$ and relaxation times $\tau_\pi,\tilde\tau_\pi$ (cf.\ \cite{Baier:2007ix})
\be
\Big{(}\tau_\pi u^\alpha \nabla_\alpha+1\Big{)}\pi^\mn=-\eta \sigma^\mn
\ee
yielding
\be\label{MISeq}
\tau_\pi \partial_\tau \phi+\frac{4}{3} \eta\, \partial_\tau \log(\y/\Z)
+[\x+\frac{4}{3} \tau_\pi \partial_\tau \log(\y^2\Z)]\phi  
=0,
\ee
and similarly for the other subsystem.
In the following we parametrize $\eta=C_\eta(\epsilon+P)/T$ and $\tau_\pi=C_\tau/T$
with dimensionless constants $C_\eta$ and $C_\tau$, where for simplicity we identify $T\equiv \epsilon^{1/4}$.
Causality requires that $C_\tau>2 C_\eta$ \cite{Romatschke:2017ejr}.

In the following we choose the first system to have ten times higher specific viscosity than the second,
$C_\eta=10\tilde C_\eta$. Having in mind subsystems that are governed by
kinetic theory
and AdS/CFT, respectively, we set \cite{Romatschke:2017ejr}
$C_\tau=5 C_\eta$ and $\tilde C_\eta=1/(4\pi),\tilde C_\tau=(2-\ln2)/(2\pi)$.
(Note that in contrast to kinetic theory, our fluid model does not
require $P_\perp$ and $P_L$ to be both positive definite.)

\begin{figure}
\includegraphics[width=.95\columnwidth]{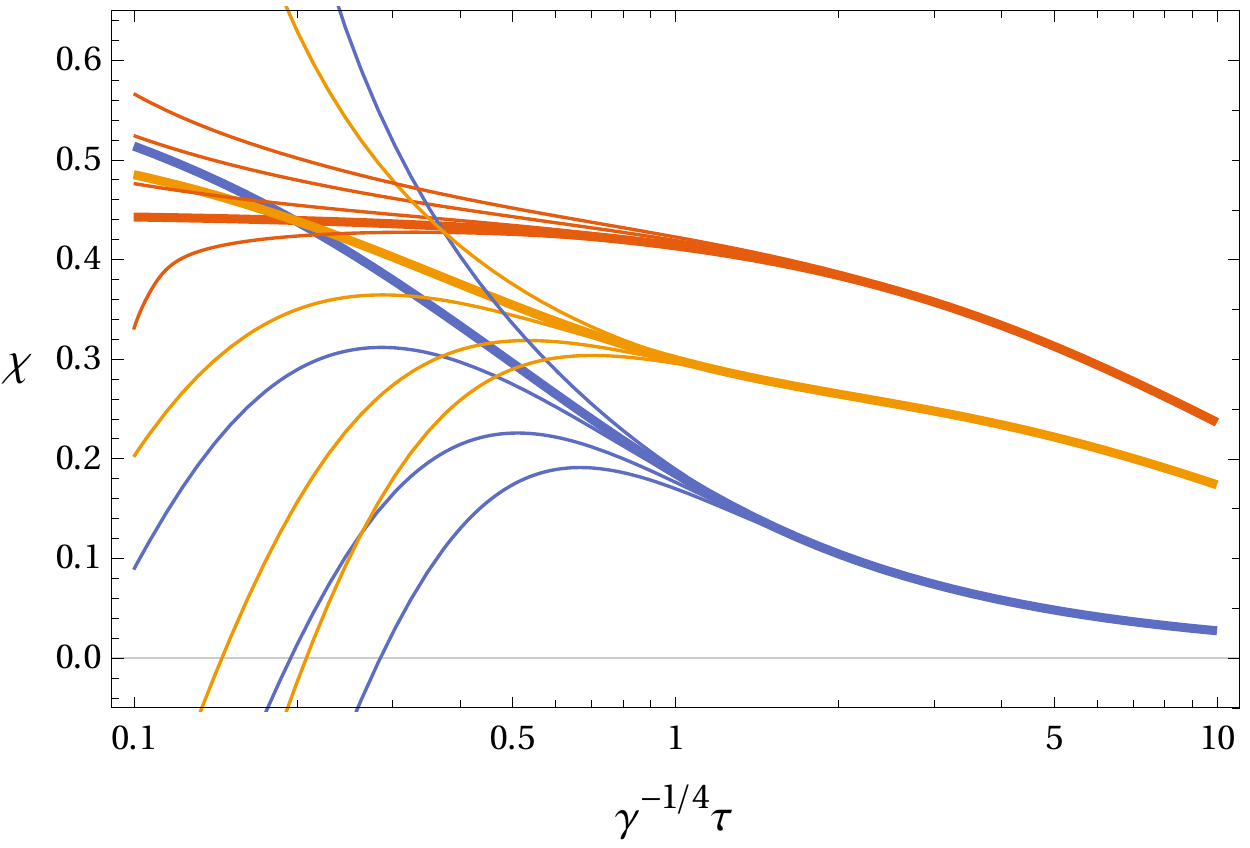}%
\caption{$\chi$'s of the attractor solution of Fig.~\ref{a11} (thick lines) and four neighboring trajectories (thin lines). 
The less viscous (strong-coupling, ``soft'') system is displayed by blue curves,
the more viscous (''hard'') system by red curves, the total system by orange curves.
\label{fig:attractor}}
\end{figure}

\begin{figure*}
\includegraphics[width=.94\columnwidth]{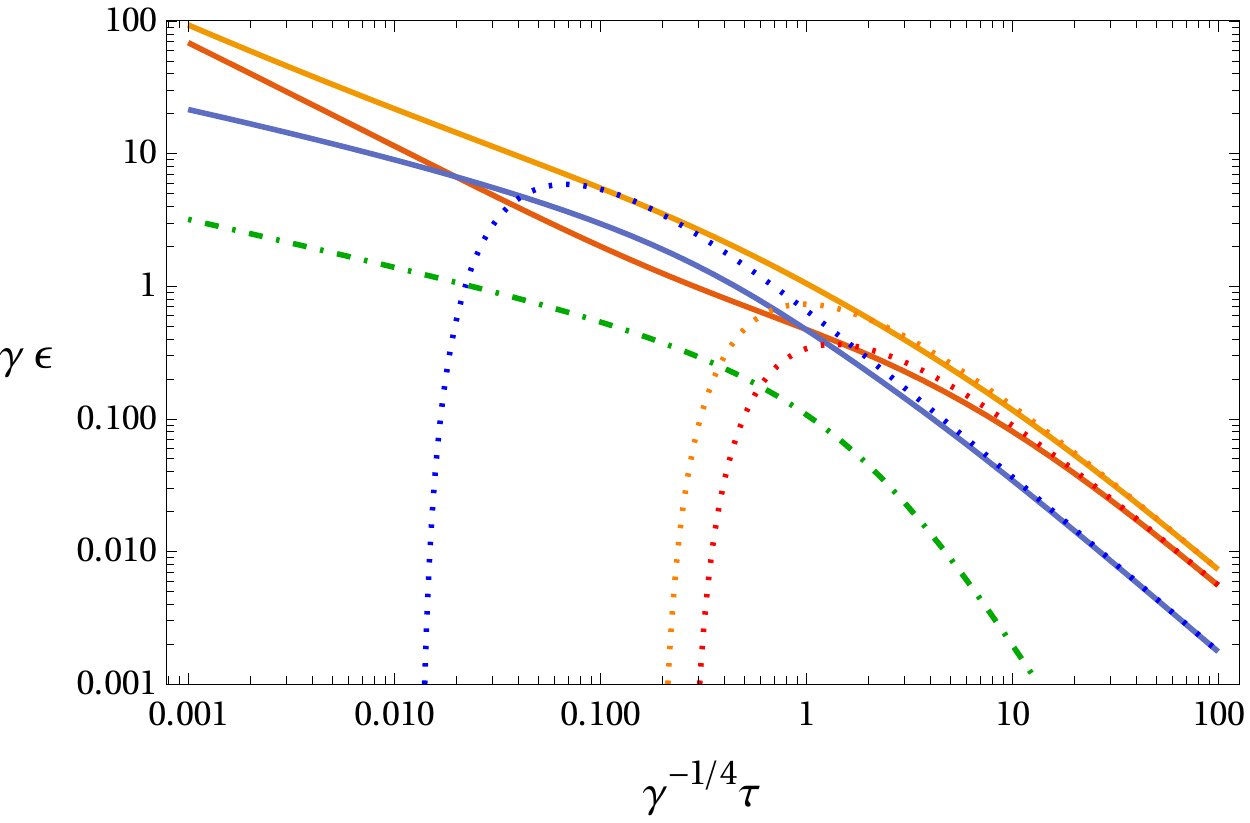}\qquad%
\includegraphics[width=.9\columnwidth]{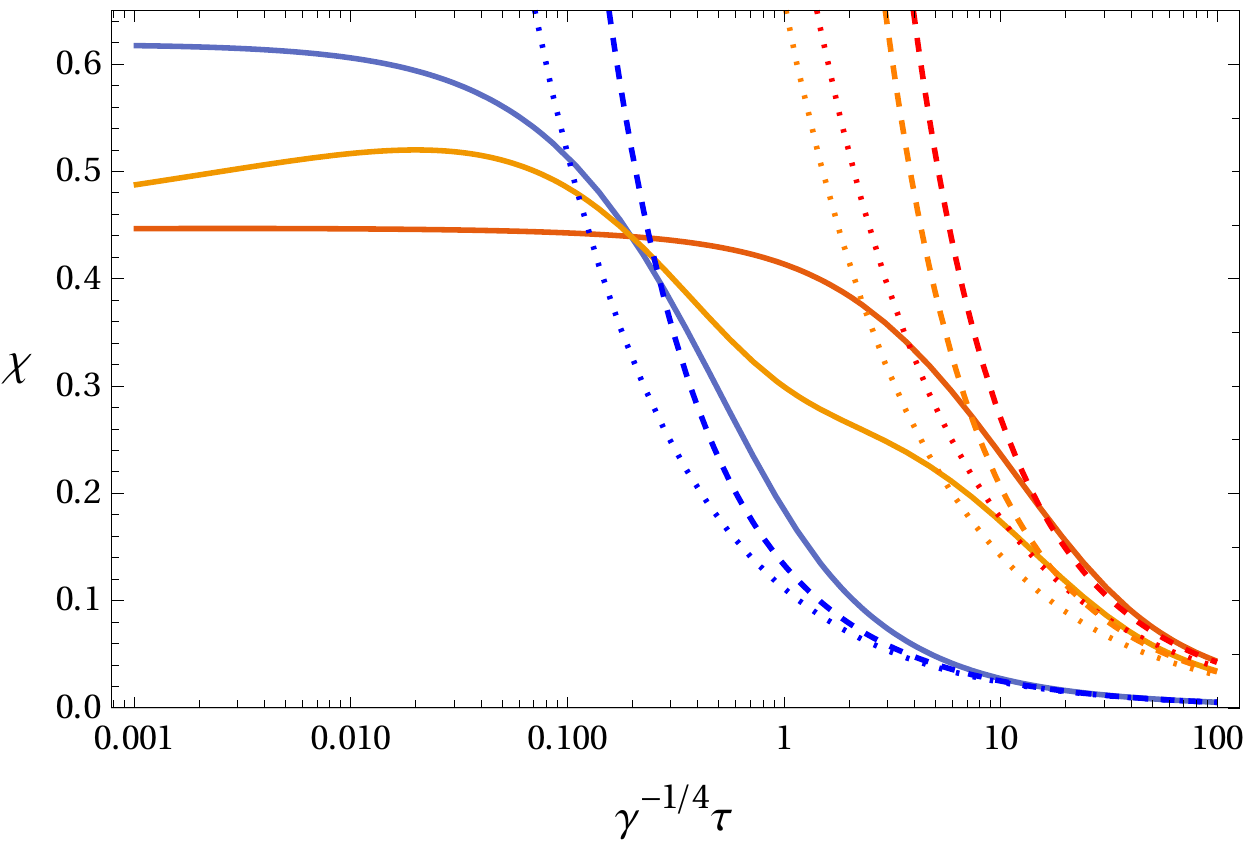}%
\caption{Energy densities and $\chi$'s of one particular attractor curve where the initially subdominant energy density
in the less viscous (``soft'') system grows larger than that in the initially dominant more viscous (``hard'') system (blue
and red curves, respectively; total system represented by orange curves).
The green dash-dotted line gives the interaction energy between the two subsystems.
In both plots, the dotted lines correspond to a first-order hydro approximation; 
in the $\chi$ plots also
the second-order hydro approximation is shown (dashed lines).
\label{a11}}
\end{figure*}

\section{Two-fluid attractor}
As found in \cite{Heller:2015dha,Denicol:2017lxn}, 
a single conformal fluid described by the two equations (\ref{WIeq}) and (\ref{MISeq}) has a hydrodynamic attractor
that can be characterized by the initial condition
\be\label{sigma}
\lim_{\tau\to0}\chi(\tau)=\sqrt{\frac{C_\eta}{C_\tau}}=:\sigma,
\ee
where\be\label{chiandA}
\chi:=\frac{\phi}{\epsilon+P}. 
\ee
(In \cite{Heller:2016rtz}
$\mathcal{A}=(P_\perp-P_L)/P$ was defined as a measure of anisotropy,
which for a conformal fluid is proportional to $\chi$: $\mathcal{A}=6\chi$.
Note also that for a conformal fluid $\chi\ge1/4$ implies $P_L\le0$.)
Generic solutions apart from the attractor solution are either singular at finite $\tau$ or 
have negative $\chi$ at early times with limiting value $-\sigma$ at $\tau=0$.
Solutions which are regular for all $\tau>0$ have positive $\epsilon$ throughout, with $\epsilon$ diverging as $\tau\to0$.

When the two fluids are coupled according to (\ref{couplingeqs}), a common attractor arises
where each fluid still has (\ref{sigma}) as limiting value, but the behavior of the energy densities
is strongly modified.
With $0<\sigma<\tilde\sigma<1/\sqrt{2}$, as is the case for our choice of parameters ($\sigma\approx 0.45$, $\tilde\sigma
\approx 0.62$), it turns out that for the common attractor 
$\epsilon$ always approaches the finite value $\sqrt{(r-1)/r}\gamma^{-1}$ and $\tilde\epsilon$ vanishes as $\tau\to0$.
But these quantities are defined with respect to the effective metric tensors $g$ and $\tilde g$ which
become singular as $\tau\to0$. Viewed from the flat Minkowski-Milne background, the contribution to
the total energy density from the first system is 
\be\mathcal{E}_1:=(a b^2 c/\tau)\epsilon\sim\tau^{4(\sigma-1)/3}
\ee 
and thus diverging for $\tau\to0$
like the single-fluid case. With $\tilde\sigma>\sigma$, the contribution of the second, less viscous
fluid is suppressed at early times as 
\be\mathcal{E}_2/\mathcal{E}_1\sim \tau^{8(\tilde\sigma-\sigma)/3}.
\ee
(Depending on the parameters $\sigma$ and $\tilde\sigma$, 
$\mathcal{E}_2$ can diverge or go to zero as $\tau\to0$; for the above
choice of parameters it also diverges, but less strongly than $\mathcal{E}_1$.)

This initial distribution of the energy densities is reminiscent of the bottom-up scenario of
thermalization in heavy-ion collisions \cite{Baier:2000sb,Kurkela:2011ub}, 
where the evolution starts out with the total energy density concentrated
in the more weakly coupled sector, which then gets redistributed to the more strongly coupled soft degrees of freedom.

Away from the limit $\tau\to0$, the differential equations and the nonlinear algebraic coupling equations
can only be solved numerically.

In Fig.~\ref{fig:attractor} we display one particular attractor solution and neighboring trajectories
in a plot of the anisotropy variables $\chi$, $\tilde\chi$, and $\chi_\mathrm{tot}$ of
the individual subsystems and the total system.

The complete set of attractor solutions is in fact a two-dimensional manifold which can be parametrized
by the dimensionless energy densities of the subsystems $\gamma\epsilon$ and $\gamma\tilde\epsilon$
at some nonzero reference time.

The attractor surface also depends on
$r\equiv-\gamma'/\gamma>1$. When $r$ is sufficiently close to 1,
the two-fluid system exhibits a first-order phase transition (with a second-order endpoint at $r=r_c$)
as analyzed in detail in \cite{Kurkela:2018dku}.
Here we have chosen $r=2$ so that there is only a cross-over behavior during the evolution of the system
as is indeed the case for QCD for high temperature and small quark chemical potential.

In Fig.~\ref{a11} the particular attractor solution in Fig.~\ref{fig:attractor} is displayed in more detail.
In the left panel, the evolution of the energy densities $\mathcal{E}_1,\mathcal{E}_2,\mathcal{E}_\mathrm{total}$
are shown for initial 
conditions which lead to $\mathcal{E}_1=\mathcal{E}_2$ at $\gamma^{-1/4}\tau=1$
and more energy in the less viscous system for a stretch of time before.

\begin{figure}[b]
\includegraphics[width=.935\columnwidth]{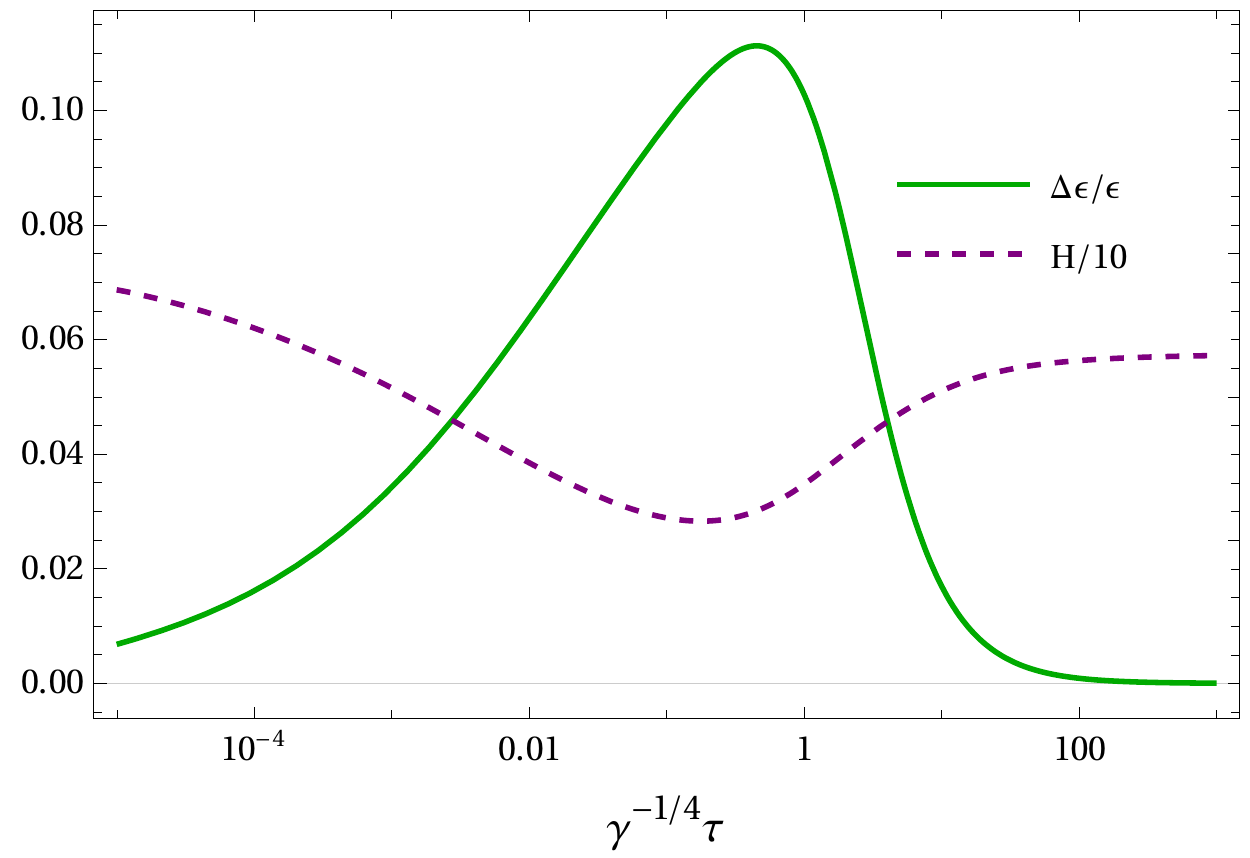}%
\caption{Interaction energy between the two subsystems over total energy (full green line), which is proportional
to the trace of the total energy-momentum tensor,
and the averaged shear viscosity
$\mathsf{H}$ (dashed purple line) as defined in (\ref{etaav})---the latter rescaled
to fit in the same plot.}\label{fig:DeltaE}
\end{figure}

Other attractor solutions exist where $\mathcal{E}_2$ exceeds $\mathcal{E}_1$ for a longer time or where
$\mathcal{E}_2$ never reaches $\mathcal{E}_1$. However, common to all is, as already explained, that
at early times the initial energy is concentrated in the more viscous (``hard'') sector which
during the Bjorken expansion is gradually transfered
to the less viscous (``soft'') sector and to the interaction energy of the two subsystems (the latter is also displayed
in Fig.~\ref{a11} and as a ratio in a linear plot in Fig.~\ref{fig:DeltaE}).
As the system cools down towards its cross-over regime, the relative amount of interaction energy rises and then switches
off rapidly.
Subsequently the two subsystems as well as the total system all approach perfect-fluid behavior
$\epsilon\sim\tau^{-4/3}$. 
Here any similarity to heavy-ion collisions of course ends, which instead ends with free-streaming of hadrons.
At larger times a large fraction of the energy is transferred back to the more weakly interacting sector.
Although this is reminiscent of the transition to a more weakly interacting hadron gas after the cross-over transition
of QCD, it is clear that our model can be considered as a toy model for heavy-ion collisions only with regard to certain aspects.
Also, the boost-invariant Bjorken flow is only relevant for its early stages.

In the right panel of Fig.~\ref{a11}, the evolution of anisotropy in the subsystems and in the full system
are displayed. While the subsystems show a monotonic decrease of anisotropy, this is not the case for the full system,
where $\chi_\mathrm{tot}$ interpolates the values in the subsystems in a more complicated manner.

From an analysis of the asymptotic behavior, which can be performed analytically and will be discussed in more detail
elsewhere, one can show that full system behaves like a single viscous fluid with effective shear viscosity
given by
\be\label{etaav}
C_\eta^\mathrm{eff}=\lim_{\tau\to\infty}\mathsf{H}(\tau)=\lim_{\tau\to\infty}
\frac{C_\eta \epsilon^{4/3}(\tau)+\tilde C_\eta \tilde\epsilon^{4/3}(\tau)}{[\epsilon(\tau)+\tilde\epsilon(\tau)]^{4/3}},
\ee
with the function $\mathsf{H}$ displayed in Fig.~\ref{fig:DeltaE} (showing a perhaps fortuitous resemblance to
the expected behavior of the shear viscosity across the transition from the deconfined phase of QCD to a hadron gas).

In both plots of Fig.~\ref{a11}, the results are compared with first-order hydrodynamics approximations;
in the plots of $\chi$'s, also the second-order approximations are shown.
This comparison allows one to define hydrodynamization times. 
In order to be close to previous definitions \cite{Heller:2016rtz,Attems:2017zam}, we require
\be
\frac{|\Delta P_L|}{P}:=
\frac{|\phi-\phi_\mathrm{1st}|}{P}<0.1 \quad\mbox{for}\;\tau>\tau_\mathrm{hd}
\ee
and similarly for the second subsystem.
For the particular solution in Fig.~\ref{a11} we obtain the ratio
$R_\mathrm{hd}:={\tau_\mathrm{hd}}/{\tilde\tau_\mathrm{hd}}\approx 5.76$.
As one might expect, the hydrodynamization time is longer for the hard sector than for the soft, more strongly
interacting sector.
In Table \ref{tab:hd}, these times are given in units of $\gamma^{1/4}$ and also converted to
the dimensionless quantities $w=\mathcal{E}_1^{1/4}\tau$, $\tilde w=\mathcal{E}_2^{1/4}\tau$.
The latter roughly scale according to $4\pi\eta/s$, albeit with
some difference in the second subsystem, reflecting the fact that at $\tau=\tilde\tau_\mathrm{hd}$
there is still an important exchange of energy with the first, not yet hydrodynamized subsystem. Also given are the values of
$\chi$ at the point of hydrodynamization, showing that the more viscuous hard sector is still strongly
anisotropic (with $P_L/P\equiv 1-4\chi\approx 0.14$) at $\tau=\tau_\mathrm{hd}$, while the soft sector
is somewhat closer to isotropy (with $\tilde P_L/\tilde P\approx 0.6$) at the earlier time $\tau=\tilde\tau_\mathrm{hd}$).

\begin{table}[b]
\caption{Subsystem hydrodynamization times and the concurrent values of the dimensionless quantities
$w=\mathcal{E}_1^{1/4}\tau$, $\tilde w=\mathcal{E}_2^{1/4}\tau$,
$\chi$, and $\tilde\chi$
for three scenarios with different values of $\mathcal{E}_1(1)=\mathcal{E}_2(1)=:\mathcal{E}(1)$,
where all dimensionful quantities are given in units of $\gamma$. The last column gives the ratio
$R_\mathrm{hd}:={\tau_\mathrm{hd}}/{\tilde\tau_\mathrm{hd}}$. \label{tab:hd}}
\begin{ruledtabular}
\begin{tabular}{c|ccc|ccc|c}
$\mathcal{E}(1)$ & $\tau_\mathrm{hd}$ & $w_\mathrm{hd}/10$ & $\chi_\mathrm{hd}$ & $\tilde\tau_\mathrm{hd}$ & $\tilde w_\mathrm{hd}$ & $\tilde\chi_\mathrm{hd}$ & $R_\mathrm{hd}$ \\
\hline
0.26 & 12.0 & 0.609 & 0.215  
     & 2.08 & 1.42 & 0.101 & 5.76 \\
0.32 & 10.2 & 0.705 & 0.203
     & 3.90 & 2.82 & 0.0525 & 2.62 \\
0.052 & 25.5 & 0.608 & 0.210
      & 1.39 & 0.613 & 0.211 & 18.4 \\
\end{tabular}
\end{ruledtabular}
\end{table}

This ordering is rather generic \footnote{We have also managed to find counterexamples, but
only in extreme choices of initial conditions where the soft sector is so suppressed at all times that it hardly
contributes to the total energy-momentum tensor.}, but the ratio $R_\mathrm{hd}$ depends strongly on the initial conditions.
In order to illustrate this, two further cases are also shown in Table \ref{tab:hd}.
When the total energy in the system at 
$\gamma^{-1/4}\tau=1$ where we have set $\mathcal{E}_1=\mathcal{E}_2$ is 
increased until we reach the limits of the attractor hypersurface, the ratio is reduced to $R_\mathrm{hd}\approx 2.62$.
Conversely, if the total energy at this reference point is made as small as possible, the ratio
becomes $R_\mathrm{hd}\approx 18.4$.
This dependence on the extra parameters of the attractor hypersurface
could perhaps be interpreted as a hint that the hydrodynamic evolution of
so-called small systems in nuclear collisions (high multiplicity events in $p$-$A$ and $p$-$p$ collisions)
\cite{Loizides:2016tew,Shen:2016zpp,Schlichting:2016sqo,Chesler:2015bba}
may involve a markedly different hydrodynamization scenario than the 
larger systems produced in heavy-ion collisions.

A more complete exploration of the possible evolutions of the hybrid fluid system presented here
as well as a study of the details how a general solution decays onto the attractor will
be the subject of a longer publication.

To conclude, we believe to have demonstrated that the hybrid fluid model introduced
in \cite{Kurkela:2018dku}, where its equilibrium properties have been analyzed, 
in combination with MIS equations
provides a novel and interesting model for the nonequilibrium dynamics of a two-component system
with different amounts of self-interactions and transport coefficients.
Clearly, many generalizations of this model can be entertained. For example, with smaller parameter $r$
one can also include first-order phase transitions, and the equation of state of the
subsystems need not be conformal.

In the special application to boost-invariant (Bjorken) flow \footnote{Recently, 
the identification of the hydrodynamic attractor using Principal Component Analysis in evolutions that are more complex than the Bjorken flow has been discussed in \cite{Heller:2020anv}. It would also be interesting
to extend this kind of studies to the case of the hybrid fluid considered here.}
considered here we have found an intriguing
model of bottom-up thermalization, which has been proposed in the context of
perturbative QCD \cite{Baier:2000sb,Kurkela:2011ub}. Remarkably, this scenario turned out to be
a universal feature of our model, if $\tilde\sigma>\sigma$, as is the case for a strongly coupled
sector with transport coefficients as given by AdS/CFT on the one hand and a weakly coupled sector with
larger specific shear viscosity on the other hand.

\section*{Acknowledgments}
AM acknowledges support from the Ramanujan Fellowship and ECR award of the Department of Science and Technology of India and also the New Faculty Seed Grant of IIT Madras.  AS is supported by the Austrian Science Fund (FWF), project no. J4406.


\bibliography{hybrid}

\end{document}